# A Systematic Literature Review of Survey Research on Engineering Identity

**Author(s):** Amdad Ahmed Awsaf, Florida International University; Heidi Cian, Maine Mathematics and Science Alliance; Remy Dou, University of Miami

## Abstract

Over the last several years, engineering identity research has gained prominence in education research circles, motivated by connections between "identity" and widely-desired outcomes such as career pursuits and selection of STEM college majors. However, quantitative research on engineering identity remains limited, with most studies focusing on qualitative research. This systematic literature review examines 85 research papers from 2007 to 2022, concentrating on survey measures related to engineering identity. Results of the review show a) a concentration of research at the postsecondary levels, with K-12 students underrepresented, and b) heavy male gender representation in survey samples. The analysis also revealed that several theoretical frameworks are frequently employed to operationalize engineering identity, but with preference for the Engineering Identity Development Scale, the Physics Identity Scale, and the Critical Engineering Agency Framework. Drawing from our analysis, we consider the consequences of both 1) existing age and gender gaps in quantitative engineering identity research and 2) preferences of particular identity frameworks within and across age and gender categories. We discuss how these possible consequences point to focal areas in engineering identity survey-based research.

## Subject/Problem

Contemporary research has regarded engineering identity as a means of better understanding student motivation in selecting engineering majors, completing their degrees, and entering the workplace (Hughes et al., 2019). According to Godwin and colleagues (2016), comprehending how individuals perceive themselves as engineers could offer nuanced understandings of student experiences and the challenges facing members of underrepresented groups in engineering. This understanding is critical for devising effective initiatives to enhance diversity and promote inclusivity within the engineering field (Hartman et al., 2019). This may also explain why, in recent years, there has been a surge in scholarly investigations regarding the concept of engineering identity (Rodriguez et al., 2018). Despite this surge, the field lacks a review of quantitative studies that focuses on how engineering identity is defined, operationalized, and instrumented across different participant populations.

The majority of studies examining engineering identity have used qualitative data and analysis techniques (e.g., Bergerson et al., 2014). However, survey-based research has advantages in identifying the relative effect and directionality of factors that contribute to identity development and related constructs. For instance, Godwin et al. (2016) proposed a model of engineering identity that situates engineering interests, recognition, and performance-competence as direct and indirect (in the case of the latter) contributors to identity. While their work is highly cited, it is also specific to populations of first-year college students, and so its generalizability to students in primary or secondary grade levels may be limited. Further, their work centers particular frameworks of engineering identity that do not capture other ways of conceptualizing identity (e.g., identity as a sense of belonging; Verdín et al., 2018).

Our aim is to synthesize which articulations of "engineering identity" have received the most attention in quantitative research to gauge the strengths of and gaps in our understanding so as to support and motivate studies with student populations who have lacked meaningful attention. We look across participant grade-level and gender bands, acknowledging both efforts to increase engineering affinities at younger ages (e.g., Patrick et al., 2021) and to address gender disparities in engineering (e.g., De Cohen & Deterding, 2009). While others, like Rodriguez and colleagues (2018), Patrick and Borrego (2016), and Morelock (2017), have conducted systematic literature reviews on engineering identity research to explore how the construct has been defined, these reviews do not focus specifically on quantitative research, nor do they extend to examine how identity is operationalized across different participant grade-levels or gender. Further, prior efforts have not attended to the extent of work addressing the distinct sets of barriers facing underrepresented engineering students such as women and K-12 students (Long & Mejia, 2016). Therefore, the present study is guided by the following research questions (RQs):

**RQ1:** What is the proportional breakdown of survey or measurement studies in engineering identity across participant grade-levels (or age bands) and gender?

**RQ2:** How do researchers operationalize engineering identity across studies as a factor of participant grade-levels (or age bands) and gender?

## Research Design

### Selection and Inclusion Criteria

To conduct our literature review, we followed the PRISMA guidelines for systematic reviews (Moher et al. 2009). Initially, we screened manuscripts based on their title and abstract.

Aligning with Simpson and Bouhafa's (2020) systematic review of STEM identity research, we included manuscripts that met the following criteria: a) full-text empirical studies, including journal articles, conference proceedings, conference proposals, and dissertations; b) written in English; c) measured engineering identity of individuals (not groups); and d) were available digitally through open or institutional access. Exclusion criteria were those that measured teacher identity or other kinds of disciplinary identity outside of "engineering" or did not carry out quantitative analysis. We selected to include studies published between 2007 and 2022 given the higher publication rate of engineering identity studies since 2007 (Rodriguez et al., 2018).

**Search Protocol**

We ran our search through Google Scholar, which Norris and colleagues (2008) note is comparable or better than alternative open-access search engines for returning qualifying articles for a systematic review. We explored search term combinations that combined "engineering", "identity", "survey", "measurement", "instrument", and "questionnaire". We evaluated the returns from various combinations and observed that two search queries yielded returns that reliably fit our inclusion criteria: a) *"engineering" AND "identity" AND ("survey" OR "instrument" OR "measurement" OR "questionnaire"),* and b) *"engineering identity" AND ("survey" OR "instrument" OR "measurement" OR "questionnaire")*. We reviewed the first 20 pages of returns for inclusion by reading abstracts to evaluate the articles against the inclusion and exclusion criteria. If reviewing an abstract was inconclusive, we reviewed the "Methods" section. We retained a total of 85 articles.

Our analysis roughly followed the approach taken by Radovic and colleagues (2018) in their review of mathematics identity studies. The first author recorded direct quotations indicating foundational theories, survey instruments, participant grade/age, and gender, which served as raw data for the analysis (Thomas & Harden, 2008). We extracted this information from the Abstract, Theoretical Framework, and Methods sections of retained publications.

To further organize this information the first author generated demographic categories. We classified gender representation based on the reported proportion of respondents who identified as male, female, non-binary, or no response; the reviewed surveys rarely accounted for individuals who identified as non-binary or other gender ($n = 5$ ). We then categorized these proportions using the following schema: “nearly equal gender composition”, “mostly female”, and “mostly male” respondents.

Following approaches taken by Simpson and Bouhafa (2020), we categorized participants into educational levels and enrollment status. Our “Elementary School” level included learners aged 5 to 10 years old, corresponding approximately to grades K-5 in the US; the “Middle School” level encompassed learners 11 to 13, corresponding to grades 6-8 in the US; and the “High School” level included learners 14 to 18, corresponding to grades 9-12 in the US. We classified individuals enrolled in an institute of higher education as “Undergraduate” or “Graduate” according to their enrollment. We did not encounter engineering identity studies with adults outside of university contexts. Additionally, while qualitative studies with populations of pre-school students exist (e.g., Pantoya et al., 2015), we did not encounter any that met our inclusion criteria of quantitative survey analysis.

## Analysis and Findings

### RQ 1: Distribution of Engineering Identity Survey Research by Grade-Levels & Gender

Our analysis indicated distinct population emphasis in engineering identity research since 2007, namely undergraduate, male students (see Table 1). Engineering identity measures for K-12 students, on the other hand, were relatively uncommon, though these few studies were more gender-balanced. We suspect that the over-representation of undergraduate studies may stem from the a) relative ease of collecting data from undergraduate students for faculty who carry out the research; b) simplicity of data collection from an adult population due to IRB requirements for children; and/or c) formal designation of "engineering" courses in undergraduate contexts compared to K-12 where students rarely can access an "engineering" class. Further, we consider the higher proportion of male participants accounted for in postsecondary engineering identity research as likely due to higher participation by male-identified students in engineering majors.

***Table 1.*** *Sampled populations represented in engineering identity studies as a factor of grade-levels and gender. Figures indicate the number of studies that addressed the indicated population and gender makeup.*

| | **~Equal gender composition** | **Mostly female students** | **Mostly male students** | **No data on gender composition** | ***Total by grade-levels*** |
|---|---|---|---|---|---|
| **Elementary** | 3 | 2 | 1 | 1 | *7* |
| **Middle** | 2 | 1 | 1 | 2 | *6* |
| **High** | 0 | 3 | 4 | 3 | *10* |
| **Undergrad** | 0 | 8 | 30 | 18 | *56* |
| **Grad** | 0 | 0 | 5 | 1 | *6* |
| ***Total by gender composition*** | *5* | *14* | *41* | *25* | *85* |

## RQ 2: Operationalizing Engineering Identity Across Participant Grade-Levels and Gender

### *Operationalization by Grade-Levels*

Of our 85 papers, approximately 31% (*n* = 27) referenced Hazari and colleagues' (2010) physics identity framework, which was employed across grade levels except Elementary School participants. This framework was most prevalently operationalized at the Undergraduate level, where it was used in 19 studies. Likewise, the Engineering Identity Measurement Scale developed by Godwin and colleagues (2013, 2016) was used across all grade levels except Elementary School, also appearing in high proportion in studies with Undergraduate participants (*n* = 17). The Science Identity Model (Carlone & Johnson, 2007) was operationalized in studies with Undergraduate ($n$ = 7) and Graduate participants ($n$ = 1).

We found that researchers have primarily used the Engineering Identity Development Scale (EIDS; Capobianco et al., 2012) to operationalize engineering identity at the Elementary School level. Five of the seven studies we identified at this grade level utilized this instrument. The "Draw an Engineer Test" (DAET; Knight & Cunningham, 2004) was only used in studies with Elementary ($n$ = 1), Middle ($n$ = 1), and High School ($n$ = 1) students, all of which also exhibited a distribution of mostly female participants. We saw the Computational Identity Framework (Kong & Lai, 2021) operationalized once in a study with Elementary School participants and once with Middle School participants.

## Operationalization by Gender Composition

We further considered the operationalization of theoretical frameworks in studies as a factor of participant gender composition (see Table 2). We identified five studies that accounted for non-binary or other gender identities. Within this set of studies, one utilized social psychology literature  to propose the Engineering Identity Scale as a means of measuring engineering identity (Brenner et al., 2014). Another applied Hazari and colleagues' (2010) framework, while a third utilized an instrument derived from the multidimensional Black identity framework (Chachra et al., 2008). The remaining employed the Engineering Identity Measurement Scale.

***Table 2.*** *Theoretically-backed survey instruments represented in engineering identity studies as a factor of gender.*

| | **~Equal gender composition** | **Mostly male students** | **Mostly female students** | **No data on gender composition** |
|---|---|---|---|---|
| Computational Identity Framework (Kong & Lei, 2021) | 2 | 0 | 0 | 0 |
| Engineering Identity Measurement Scale (Godwin, 2016) | 1 | 15 | 2 | 7 |
| Physics Identity Scale (Hazari et al., 2010) | 0 | 11 | 4 | 12 |
| Engineering Identity Development Scale (Capobianco et al., 2012) | 2 | 1 | 1 | 1 |
| Critical Engineering Agency Framework (Godwin et al., 2013) | 0 | 8 | 1 | 7 |
| Science Identity Model (Carlone & Johnson, 2007) | 0 | 2 | 2 | 1 |
| DAET (Knight & Cunningham, 2004) | 0 | 0 | 3 | 0 |

## Contributions & Implications

Our findings extend reviews of engineering identity studies presented by Rodriguez et al. (2018), Patrick and Borrego (2016), and Morelock (2017) by focusing on quantitative measures of engineering identity and the characteristics of participating student populations. This study sheds light on our limited understanding of engineering identity development within younger, student populations. This gap may have consequences for the design of interventions and programs aimed to stimulate children's interest and involvement in engineering (Capobianco et al., 2015). When such programs must rely on patterns derived from older students, key developmental factors may be missed in youth's identity formation.

We interpret the gender-skewed data for engineering identity studies as a symptom of inequitable access to postsecondary engineering education by learners who do not identify as male (Godwin et al., 2016). Given the connection between college major choice and identity (e.g., Godwin et al., 2013), our work suggests a cyclical reproduction: If definitions of engineering identity and related instruments are derived from male-majority population samples, perhaps the factors that are determined to support engineering identity are themselves masculine. Thus, efforts to boost engineering identity based on quantitative research may reflect on the extent to which they boost *masculine* engineering identity, specifically. Until more gender-representative research on engineering undergraduates can be conducted, such identity-based efforts may support their equity goals by purposefully considering frameworks of feminism in STEM (e.g., Heybach & Pickup, 2017).

These findings may benefit NARST members who are concerned with the consequences of and solutions to gender disparities in engineering fields and engineering education, and particularly those whose research studies involve quantitative analysis of surveys or other engineering identity measurement instruments.